\documentstyle [12pt,aaspp4]{article}
\newcommand{\simgt}{\lower.5ex\hbox{$\; \buildrel > \over \sim \;$}}
\newcommand{\simlt}{\lower.5ex\hbox{$\; \buildrel < \over \sim \;$}}

\newcommand{\citet}[1] {\cite{#1}}
\newcommand{\citep}[1] {(\cite{#1})}
\slugcomment{RESCEU-32/99\quad UTAP-343/99}
\begin{document}
\title{Density Profiles of Dark Matter Halo are not Universal}

\author{Y.P. Jing and Yasushi Suto}
\affil{Department of Physics and Research Center for
    the Early Universe (RESCEU) \\ School of Science, University of
    Tokyo, Tokyo 113-0033, Japan.}
\affil{jing@utap.phys.s.u-tokyo.ac.jp, suto@phys.s.u-tokyo.ac.jp}

\received{1999 September 7}
\accepted{1999 ???}

\begin{abstract}
  We perform a series of high -- resolution N-body simulations
  designed to examine the density profiles of dark matter halos. From
  12 simulated halos ranging the mass of $2\times10^{12}\sim
  5\times10^{14} h^{-1}{\rm M_\odot}$ (represented by $\sim 1$ million
  particles within the virial radius), we find a clear systematic
  correlation between the halo mass and the slope of the density
  profile at 1\% of the virial radius, in addition to the variations
  of the slope among halos of the similar mass. More specifically, the
  slope is $\sim -1.5$, $-1.3$, and $-1.1$ for galaxy, group, and
  cluster mass halos, respectively. While we confirm the earlier
  simulation results that the inner slope is steeper than the {\it
    universal} profile originally proposed by Navarro, Frenk \& White,
  this mass dependence is inconsistent with the several analytical
  arguments attempting to link the inner slope with the primordial
  index of the fluctuation spectrum.  Thus we conclude that the dark
  matter density profiles, especially in the inner region, are not
  universal.
\end{abstract} 


\section{Introduction}

Can one recover (some aspects of) the initial conditions of the
universe from the distribution of galaxies at $z \sim 0$ ?  A
conventional answer to this question is affirmative, {\it provided}
that the effect of a spatial bias is well understood and/or if it does
not significantly alter the interpretation of the observed
distribution.  This consensus underlies the tremendous effort in the
past and at present to extract the cosmological implications from the
existing and future galaxy redshift surveys. The two-point correlation
function $\xi(r)$ is a good example supporting this idea; on large
scales it is trivially related to the primordial spectrum of mass
fluctuations, $P_{\rm i}(k)$. Furthermore the effective power-law
index of the two-point correlation function on sufficiently small
scales is related to the initial power-law index $n_{\rm i}$ of
$P_{\rm i}(k) \propto k^{n_{\rm i}}$ as $\xi(r) \propto r^{-3(3+n_{\rm
    i})/(5+n_{\rm i})}$ (e.g. \cite{Peebles1980};
\cite{Suginohara1991}; \cite{Suto1993}).  In other words, the initial
conditions of the universe are imprinted in the behavior of galaxies
on small scales (again apart from the effect of bias).  This is why
the phenomenological fitting formulae for the nonlinear power spectrum
\citep{Hamilton1991,JMW1995,PD1996,Ma1998} turn out to be so
successful.  This fact, however, seems to be in conflict with the
universal density profile proposed by Navarro, Frenk \& White
(1996,1997; hereafter NFW) for virialized dark matter halos. In their
study, NFW selected halos which look to be virialized, and found that
the density profiles universally obey the NFW form $\rho(r)\propto
r^{-1}(r+r_s)^{-2}$.  It is yet unclear to which degree their results
are affected by their selection criterion which is not
well-defined. In general, different halos should have experienced
different merging histories depending on their environment and mass.
Thus even if the halos do have a {\it universal} density profile {\it
  statistically} (i.e., after averaging over many realizations), it is
also natural that individual halo profiles are intrinsically scattered
around the universal profile \citep{Jing1999}. Definitely this is a
matter of semantics to a certain extent; the most important finding of
NFW is that such halo-to-halo variations are surprisingly small.

A universal density profile was also reported by \cite{Moore1999} on
the basis of high -- resolution simulations of one cluster-mass halo
and four galaxy-mass halos, and they claim that the density profile
$\rho(r)\propto r^{-1.5}$ in the most inner region.  In what follows,
we will address the following quantitative and specific questions
concerning the halo profile, especially its most inner region, using
the high-resolution $N$-body simulations; the inner slope of the halo
profile is really described by $\rho(r) \propto r^{-1}$ or $\rho(r)
\propto r^{-1.5}$ {\it universally} as NFW and \cite{Moore1999}
claimed ?  If not, does the slope vary among the different halos ?  Is
there any systematic correlation between the slope and the mass of
halos ?

In fact, some of the above questions have been partially addressed
previously with different approaches and methodologies
\citep{Fukushige1997,Evans1997,Moore1998,Syer1998,Nusser1999,Jing1999,AFKK1999}.  In order to revisit those in a more systematic and
unambiguous manner, we have developed a nested grid P$^3$M N-body code
designed to the current problem so as to ensure the required numerical
resolution in the available computer resources.  This enables us to
simulate 12 realizations of halos in a low-density cold dark matter
(LCDM) universe with $(0.5 - 1)\times 10^6$ particles in a range of
mass $10^{12 \sim 15}M_\odot$.

\section{Simulation procedure}

As \citet{Fukushige1997} and later \citet{Moore1998} demonstrated, the
inner profile of dark matter halos is substantially affected by the
mass resolution of simulations. To ensure the required resolution (at
least comparable to theirs), we adopt the following two-step procedure.
A detailed description of the implementation and resolution test will
be presented elsewhere.

First we select dark matter halos from our previous cosmological
P$^3$M N-body simulations with $256^3$ particles in a $(100h^{-1}{\rm
  Mpc})^3$ cube \citep{JS1998,Jing1998}.  To be specific, we use one
simulation of the LCDM model of $\Omega_0=0.3$, $\lambda_0=0.7$,
$h=0.7$ and $\sigma_8=1.0$ according to \citet{Kitayama1997}.  The
mass of the individual particle in this simulation is
$7\times10^{9}M_\odot$.  The candidate halo catalog is created using
the friend-of-friend grouping algorithm with the bonding length of 0.2
times the mean particle separation.

We choose twelve halos in total from the candidate catalog so that
they have mass scales of clusters, groups, and galaxies (Table
\ref{table:halos}). Except for the mass range, the selection is
random, but we had to exclude about 40\% halos of galactic mass from
the original candidates since they have a neighboring halo with a much
larger mass.  We use the multiple mass method to re-simulate them.  To
minimize the contamination of the coarse particles on the halo
properties within the virial radius at $z=0$, $r_{\rm vir}$, we trace
back the particles within $3r_{\rm vir}$ of each halo to their initial
conditions at redshift $z=72$.  This is more conservative than that
adopted in previous studies, and in fact turned out to be important
for galactic mass halos.  Note that we define $r_{\rm vir}$ such that
the spherical overdensity inside is $\sim 18\pi^2 \Omega_0^{0.4} \sim
110$ times the critical density, $\rho_{\rm crit}(z=0)$.

Then we regenerate the initial distribution in the cubic volume
enclosing these halo particles with larger number of particles by
adding shorter wavelength perturbation to the identical initial
fluctuation of the cosmological simulation.  Next we group fine
particles into coarse particles (consisting of at most 8 fine
particles) within the high-resolution region if they are not expected
to enter the central halo region within $3r_{\rm vir}$. As a result,
there are typically $2.2\times 10^6$ simulation particles, $\sim
1.5\times 10^6$ fine particles and $\sim 0.7\times 10^6$ coarse
particles for each halo.  Finally about $(0.5 - 1)\times 10^6$
particles end up within $r_{\rm vir}$ of each halo. Note that this
number is significantly larger than those of NFW, and comparable to
those of \citet{Fukushige1997} and \citet{Moore1998,Moore1999}. The
contamination of the coarse particles, measured by the ratio of the
mass of the coarse particles within the virial radius to the total
virial mass, is small, about $10^{-4}$, $10^{-3}$, and $10^{-2}$ for
cluster, group, and galactic halos respectively.

We evolve the initial condition for the selected halo generated as
above using a new code developed specifically for the present purpose.
The code implements the nested-grid refinement feature in the original
P$^3$M N-body code of \citet{JF1994}.  Our code implements a constant
gravitational softening length in comoving coordinates, and we change
its value at $z=4$, 3, 2, and 1 so that the proper softening length
(about 3 times the Plummer softening length) becomes 0.004 $r_{vir}$.
Thus our simulations effectively employ the constant softening length
in proper coordinates at $z \le 4$.  The first refinement is placed to
include all fine particles, and the particle-particle (PP) short range
force is added to compensate for the larger softening of the
particle-mesh (PM) force. When the CPU time of the PP computation
exceeds twice the PM calculation as the clustering develops, a second
refinement is placed around the center of the halo with the physical
size about $1/3$ of that of the first refinement.  The mesh size is
fixed to $360^3$ for the parent periodic mesh and for the two isolated
refinements. The CPU time for each step is about $1.5$ minutes at the
beginning and increases to $5$ minutes at the final epoch of the
simulation on one vector processor of Fujitsu VPP300 (peak CPU speed
of 1.6 GFLOPS). A typical run of $10^4$ time steps, which satisfies
the stability criteria (Couchman et al. 1995), takes $700$ CPU hours to
complete.

\section{Results}

\begin{figure}
\epsscale{1.0} \plotone{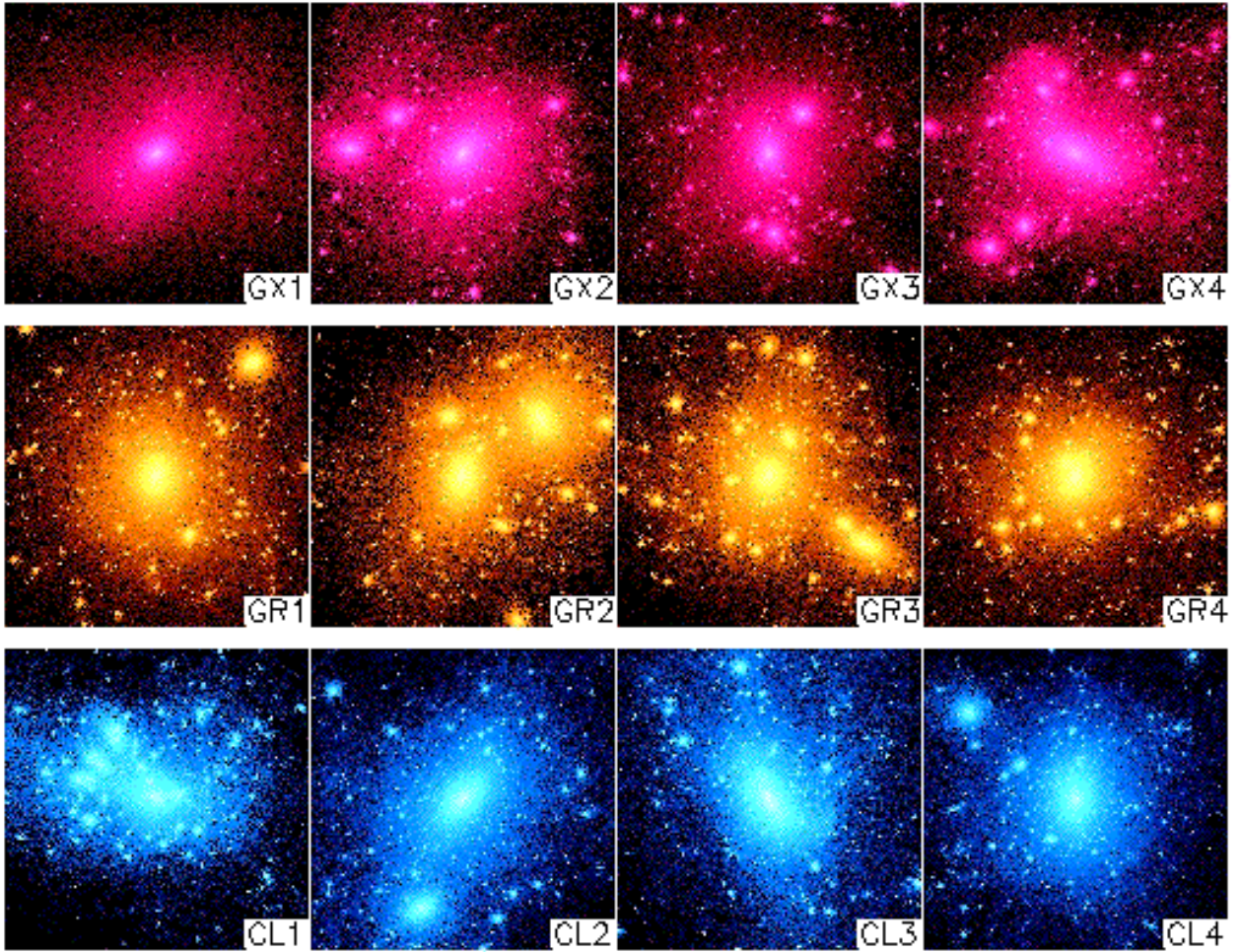}
\caption{Snapshots of the simulated halos at $z=0$.  Left, middle and
  right panels display the halos of galaxy, group and cluster masses,
  respectively (see Table 1). The size of each panel corresponds to
  $2r_{\rm vir}$ of each halo.  \label{fig:haloplot} }
\end{figure}

\begin{figure}
\epsscale{1.0} \plotone{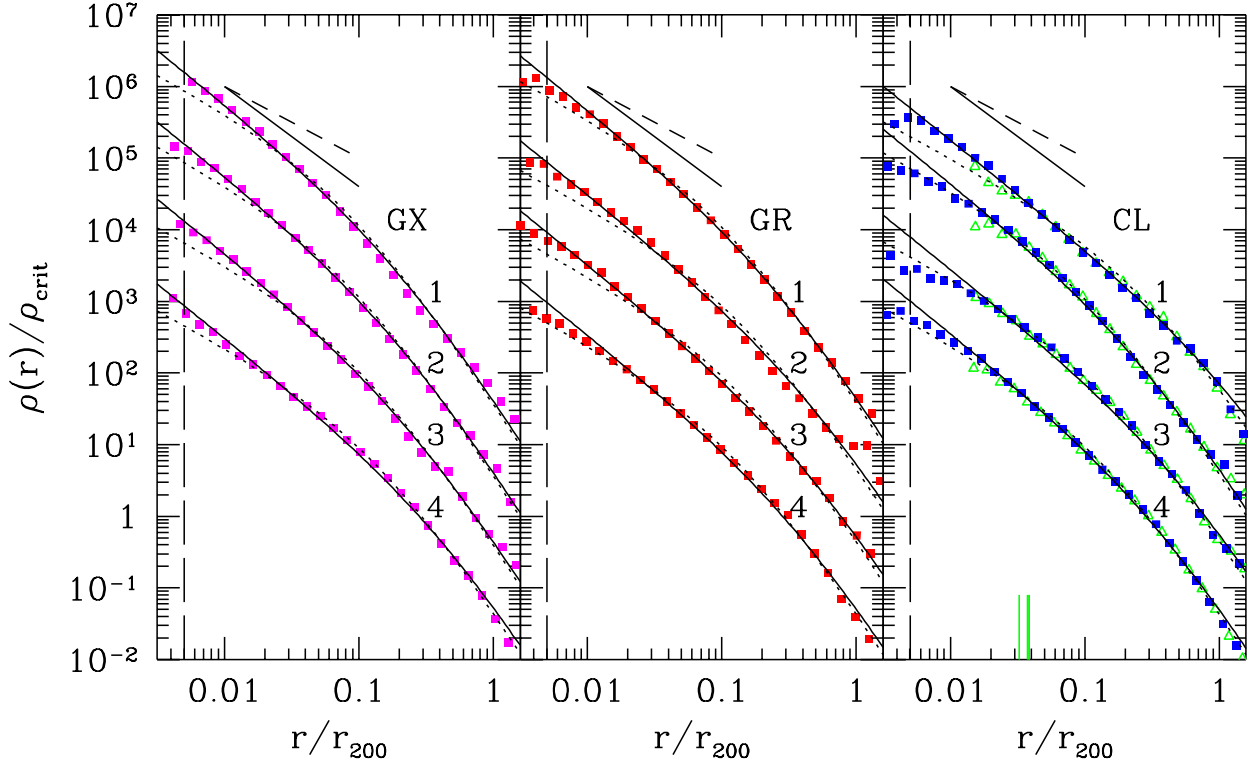}
\caption{Spherically-averaged radial density profiles of the simulated
halos of galaxy ({\it left}), group ({\it middle}), and cluster ({\it
right}) masses.  The solid and dotted curves represent fits of
$\beta=1.5$ and $\beta=1$ respectively (see text for the fit form).
For reference, we also show $\rho(r) \propto r^{-1}$ and $r^{-1.5}$ in
dashed and solid lines.  The vertical dashed lines indicate the force
softening length which corresponds to our resolution limit. The open
triangles in the right panel show the results for the corresponding
halos in the lower-resolution cosmological simulation, and the long
ticks at the bottom mark the force softening of the cosmological
simulation.  For the illustrative purpose, the values of the halo
densities are multiplied by 1, $10^{-1}$, $10^{-2}$, $10^{-3}$ from
top to bottom in each panel.  \label{fig:haloprof} }
\end{figure}

\begin{figure}
\epsscale{1.0} \plotone{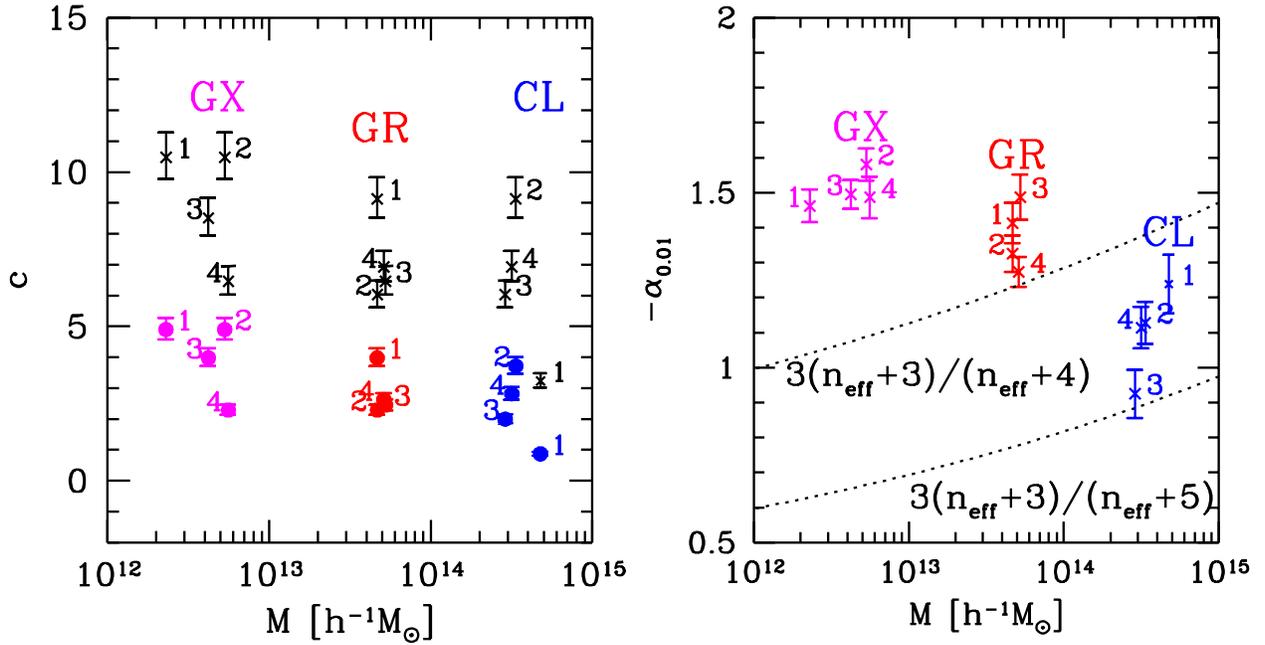}
\caption{{\it Left panel:} the concentration parameters for each
  halo for the Moore et al. (1999) form (filled circles) and for the
  NFW form (crosses). Numbers labeling each symbol correspond to the
  halo ID in Table 1. {\it Right panel:} Power-law index of the inner
  region ($0.007 <r/r_{200}<0.02$) as a function of the halo mass. The
  upper and lower dotted curves indicate the predictions of Hoffman \&
  Shaham (1985) and Syer \& White (1996), respectively.
\label{fig:haloslope}}
\end{figure}

Figure \ref{fig:haloplot} displays the snapshot of the twelve halos at
$z=0$.  Clearly all the halos are far from spherically symmetric, and
surrounded by many substructures and merging clumps.  This is
qualitatively similar to that found by \citet{Moore1998,Moore1999} for
their high-resolution halos in the $\Omega_0=1$ CDM model.  The
corresponding radial density profiles are plotted in Figure
\ref{fig:haloprof}. The halo center is defined as the position of the
particle which possesses the minimum potential among the particles
within the sphere of radius $r_{vir}$ around the center-of-mass of the
fine particles.  In spite of the existence of apparent sub-clumps
(Fig.  \ref{fig:haloplot}), the spherically averaged profiles are
quite smooth and similar to each other as first pointed out by NFW.
The inner slope of the profiles, however, is generally steeper than
the NFW value, $-1$, in agreement with the previous findings of
\citet{Fukushige1997} and \cite{Moore1998}.  We have fitted the
profiles to $\rho(r)\propto r^{-\beta}(r+r_s)^{-3+\beta}$ with
$\beta=1.5$ (similar to that used by Moore et al.1999; the solid
curves) and $\beta=1$ (NFW form; the dotted curves) for
$0.01r_{200}\le r \le r_{200}$, where $r_{200}$ is the radius within
which the spherical overdensity is $200 \rho_{\rm crit}(z=0)$.  The
resulting concentration parameter $c$, defined as the $r_{200}/r_s$,
is plotted in the left panel of Figure \ref{fig:haloslope}. This is
the most accurate determination of the concentration parameter for the
LCDM model. There exists a significant scatter among $c$ for similar
mass \citep{Jing1999}, and a clear systematic dependence on halo mass
(NFW, \citet{Moore1999}).

The most important result is that the density profiles of the 4
galactic halos are all well fitted by $\beta=1.5$, but those of the
cluster halos are better fitted to the NFW form $\beta=1$. This is in
contrast with \citet{Moore1999} who concluded that both galactic and
cluster halos have the inner density profile $\rho(r)\propto
r^{-1.5}$, despite that they considered one cluster-mass halo
alone. In fact, our current samples can address this question in a
more statistical manner. CL1 has significant substructures, and the
other three are nearly in equilibrium. Interestingly the density
profiles of CL2 and CL3 are better fitted to the NFW form, and that of
CL4 is in between the two forms. The density profiles of the group
halos are in between the galactic and cluster halos, as expected. One
is better fitted to the NFW form, whereas the other three follow the
$\beta=1.5$ form.

To examine this more quantitatively, we plot the inner slope fitted to
a power-law for $0.007 <r/r_{200}<0.02$ as a function of the halo mass
in the right panel of Figure \ref{fig:haloslope}. This figure
indicates two important features; a significant scatter of the inner
slope among the halos with similar masses and a clear systematic trend
of the steeper profile for the smaller mass.  For reference we plot
the predictions for the slope, $-3(3+n)/(4+n)$ by \citet{Hoffman1985}
and $-3(3+n)/(5+n)$ by \citet{Syer1998}, using for $n$ the effective
power-law index $n_{\rm eff}$ of the linear power spectrum at the
corresponding mass scale(Table 1). With a completely different
methodology, \citet{Nusser1999} argue that the slope of the density
profile within $\sim 0.01r_{\rm vir}$ is in between the above two
values.  On the basis of the slope -- mass relation that we
discovered, we disagree with their interpretation; for the galactic
halos, the analytical predictions could be brought into agreement with
our simulation only if the effective slope were $-2$, which is much
larger than the actual value $-2.5$ on the scale.

We would like to emphasize that our results are robust against the
numerical resolution for the following reasons. Since we have used the
same time steps and the same force softening length in terms of
$r_{200}$, the resolution effect, which is generally expected to make
the inner slope of $\rho(r)$ shallower, should influence the result of
the galactic halos more than that of cluster halos. In fact this is
opposite to what we found in the simulation. Furthermore our
high--resolution simulation results agree very well with those of the
lower--resolution cosmological simulations (open triangles) for the
cluster halos on scales larger than their force softening length (short
thin lines at the bottom of Figure \ref{fig:haloprof}).  We have also
repeated the simulations of several halos employing 8 times less
particles and 2 times larger softening length, and made sure that the
force softening length $\sim 0.005 r_{200}$ (the vertical dashed lines
of Figure \ref{fig:haloprof}) is a good indicator for the resolution
limit.

\section{Conclusion and Discussion}

In this {\it Letter} we have presented the results of the largest,
systematic study on the dark matter density profiles. This is the
first study which simulates a dozen of dark halos with about a million
particles in a flat low-density CDM universe.  This enables us to
address the profile of the halos with unprecedented accuracy and
statistical reliability.  While qualitative aspects of our results are
not inconsistent with those reported by Moore et al. (1999), our
larger sample of halos provides convincing evidence that {\it the
form} of the density profiles is not universal; instead it depends on
halo mass. Since mass and formation epoch are linked in hierarchical
models, the mass dependence may reflect an underlying link to the age
of the halo. Older galactic halos more closely follow the $\beta=1.5$
form while younger cluster halos have shallower inner density profiles
fitted better by the NFW form. Whether this difference represents secular
evolution remains to be investigated in future experiments.

Our results are not fully expected by the existing analytical
work. Although the analytical work
\citep{Syer1998,Nusser1999,Lokas1999} concluded that the inner profile
should be steeper than $-1$, their interpretation and/or predicted
mass-dependence are different from our numerical results. This implies
that while their arguments may cover some parts of the physical
effects, they do not fully account for the intrinsically complicated
nonlinear dynamical evolution of non-spherical self-gravitating
systems.

We also note that the small-scale power which was missed in the
original cosmological simulation has been added to the initial
fluctuation of the halos. The fact that each halo has approximately
the same number of particles means that more (smaller-scale) power has
been added to the low mass halos than to the high mass ones. It is yet
unclear how much effect this numerical systematics would have on the
mass dependence of the inner slope found in this paper, and we will
investigate this question in future work.
  
In summary, the mass dependence of the inner profile indicates the
difficulty in understanding the halo density profile from the
cosmological initial conditions in a straightforward manner.  Even if
the density profiles of dark halos are not universal to the extent
which NFW claimed, however, they definitely deserve further
investigation from both numerical and analytical points of view.

\acknowledgments

We thank J. Makino for many stimulating discussions and suggestions,
and the referee for a very detailed report which significantly
improves the presentation of this paper.  Y.P.J. gratefully
acknowledges support from a JSPS (Japan Society for the Promotion of
Science) fellowship.  Numerical computations were carried out on
VPP300/16R and VX/4R at ADAC (the Astronomical Data Analysis Center)
of the National Astronomical Observatory, Japan, as well as at RESCEU
(Research Center for the Early Universe, University of Tokyo) and KEK
(High Energy Accelerator Research Organization, Japan). This research
was supported in part by the Grant-in-Aid by the Ministry of
Education, Science, Sports and Culture of Japan (07CE2002) to RESCEU,
and by the Supercomputer Project (No.99-52) of KEK.



\newpage

\begin{deluxetable}{cccccccccccccc}
\tablecaption{Properties of the simulated halos}
\tablewidth{0pt}
\tablehead{
\colhead{ID}
&\colhead{$M$ \tablenotemark{a}}
&\colhead{$N_p$ \tablenotemark{b}}
&\colhead{$r_{\rm vir}$\tablenotemark{c}}
&\colhead{$n_{\rm eff}$ \tablenotemark{d}} }
\startdata
GX 1 & $2.30\times 10^{12}$ & 458,440 & 0.269& $-2.47$\cr
GX 2 & $5.31\times 10^{12}$ & 840,244 & 0.356& $-2.42$\cr
GX 3 & $4.21\times 10^{12}$ & 694,211 & 0.330& $-2.44$\cr
GX 4 & $5.60\times 10^{12}$ & 1,029,895 & 0.363& $-2.42$\cr
GR 1 & $4.66\times 10^{13}$ & 772,504 & 0.735& $-2.31$\cr
GR 2 & $4.68\times 10^{13}$ & 907,489 & 0.736& $-2.31$\cr
GR 3 & $5.24\times 10^{13}$ & 831,429 & 0.764& $-2.30$\cr
GR 4 & $5.12\times 10^{13}$ & 901,518 & 0.758& $-2.30$\cr
CL 1 & $4.77\times 10^{14}$ & 522,573 & 1.59& $-2.12$\cr
CL 2 & $3.36\times 10^{14}$ & 839,901 & 1.42& $-2.14$\cr
CL 3 & $2.89\times 10^{14}$ & 664,240 & 1.35& $-2.16$\cr
CL 4 & $3.17\times 10^{14}$ & 898,782 & 1.39& $-2.15$\cr
\enddata
\small{
\tablenotetext{a}{Mass of the halo within its virial radius, in units
of $h^{-1}{\rm M_\odot}$.}
\tablenotetext{b}{Number of particles within its virial radius.}
\tablenotetext{c}{the virial radius, in units of $h^{-1}$Mpc.}
\tablenotetext{d}{the effective slope of the linear power spectrum at
the halo mass scale.}
}
\label{table:halos}
\end{deluxetable}

\end{document}